\documentclass[conference]{IEEEtran}
\IEEEoverridecommandlockouts
\usepackage{cite}
\usepackage{amsmath,amssymb,amsfonts}
\usepackage{algorithmic}
\usepackage{graphicx}
\usepackage{textcomp}
\usepackage{xcolor}

\def\BibTeX{{\rm B\kern-.05em{\sc i\kern-.025em b}\kern-.08em
    T\kern-.1667em\lower.7ex\hbox{E}\kern-.125emX}}
\begin{document}

% \title{Joint Optimization of Buffer Delay and Hybrid Automatic Repeat Request for Video Communications}\\
\title{Joint Optimization of Buffer Delay and HARQ for Video Communications}

\author{\IEEEauthorblockN{Baoping Cheng\IEEEauthorrefmark{1,2},
Peng Lei\IEEEauthorrefmark{2},
Xiaoyan Xie\IEEEauthorrefmark{2},
Tao Fu\IEEEauthorrefmark{2},
Yukun Zhang\IEEEauthorrefmark{2},and
Xiaoming Tao\IEEEauthorrefmark{1},}
\IEEEauthorblockA{\IEEEauthorrefmark{1}Department of Electronic Engineering, Tsinghua University, Beijing, China}
\IEEEauthorblockA{\IEEEauthorrefmark{2}China Mobile (Hangzhou) Information Technology Co., Ltd, Hangzhou, China}
\IEEEauthorblockA{Email:cbp21@mails.tsinghua.edu.cn, taoxm@tsinghua.edu.cn}
\IEEEauthorblockA{\{leipeng, xiexiaoyan, futao, zhangyukun\}@cmhi.chinamobile.com}
}

\maketitle

\begin{abstract}
To improve the quality of experience (QoE) in video communication over lossy networks, this paper presents a transmission method that jointly optimizes buffer delay and Hybrid Automatic Repeat request (HARQ), referred to as BD-HARQ. This method operates on packet group and employs dynamic buffer delay combined with HARQ strategy for transmission. By defining the QoE based on metrics such as buffer delay, Forward Error Correction (FEC) redundancy, and data recovery rate, the proposed method  derives its closed-form expression through rigorous mathematical modeling and analysis. The optimal transmission parameters, i.e., the buffer delay and the FEC redundancy, are then determined and implemented, guaranteeing the real-time performance, transmission efficiency, and data recovery rate of video communication.  Experimental results demonstrate that the proposed method aligns well with its theoretical expectations, and  that it can provide up to 13.7\% higher QoE compared to existing methods and increase the tolerance for packet loss rate from 15\%-22\% to up to 31\% while maintaining a high QoE.

\end{abstract}

\begin{IEEEkeywords}
video communications, buffer delay, hybrid automatic repeat request, quality of experience
\end{IEEEkeywords}

\section{Introduction}

Video communication has become integral to our daily lives and work. However, with the increasing demand for real-time transmission, issues such as unstable wireless signals, fluctuating bandwidth, and network congestion lead to frequent packet loss, significantly affecting the Quality of Experience (QoE) for users. Video semantic communication technology \cite{qin2021semantic}, which mimics human visual perception, can effectively reduce the bitrate of transmitting videos, showing great potential. Nevertheless, improving system QoE in lossy networks remains a substantial challenge.

In traditional video communication, extensive research has been conducted on enhancing QoE performance in lossy networks, employing techniques such as Forward Error Correction (FEC)\cite{huo2015tutorial} , Hybrid Automatic Repeat Request (HARQ) \cite{Ahmed2021Hybrid}, and receiving buffer delay \cite{holmer2013handling}. Xiao et al. \cite{xiao2013real} proposed a real-time video streaming solution that employs randomly expanded Reed-Solomon (RS) code for FEC, enhancing error correction performance with a certain decoding delay by including both current and previous frame video packets in RS code blocks. Cheng et al. \cite{cheng2020deeprs} introduced a deep learning-based network-adaptive FEC method, DeepRS, which uses deep learning to predict packet loss patterns in real-time video streams, enabling more appropriate FEC redundancy allocation and improving packet loss recovery. Holmer et al. \cite{holmer2013handling} explored packet loss handling mechanisms in Web Real-Time Communications (WebRTC) and proposed an adaptive HARQ method, which adjusts receiver buffer delay and sender redundancy based on the statistical interval between consecutive frames at the receiver, balancing video quality, playback jitter, and delay. Rapaport et al.  \cite{rapaport2013adaptive} proposed a layered video transmission method, which prioritizes video packets based on their importance and executes additional HARQ retransmissions for high-priority video packets to reduce the impact of packet loss, thereby improving overall transmission quality. Despite these advancements in traditional video communication, there remains a lack of research on the joint optimization of dynamic buffer delay and HARQ, particularly with mathematical modeling and analysis combined with network state considerations. 

In video semantic communication, recent studies have also been aimed to improve video performance under constrained networks. Huang et al. \cite{huang2021deep} utilized a Generative Adversarial Network (GAN) based image semantic codec to enhance semantic consistency in received images and reduce transmission data volume. Gong et al. \cite{gong2023adaptive} proposed an adaptive bitrate video semantic communication system, combining a Swin Transformer-based semantic codec with an actor-critic-based adaptive bitrate algorithm to improve robustness against network fluctuations and enhance the accuracy and real-time performance of classification tasks. Zhang et al. \cite{zhang2024improving} introduced a deep lossy transmission paradigm based on semantic communication, achieving end-to-end joint source-channel coding through a deep video semantic encoding model, leveraging its data compression and error correction capabilities to improve video frame recovery quality. Li et al. \cite{li2024video} treated target objects as critical information, reducing data volume through primary object extraction and context encoding, thereby improving video transmission efficiency. However, these studies primarily focus on exploiting the characteristics of semantic communication tasks for targeted semantic extraction and encoding to reduce bit rate or enhance decoding recovery capabilities, without considering the optimization of transmission using traditional strategies such as dynamic buffer delay and HARQ, which to some extent limits the system’s QoE. 
 
To further improve the QoE in video communication over lossy networks, this paper introduces a transmission method that jointly optimizes dynamic buffer delay and HARQ, referred to as BD-HARQ. The main contributions of this paper are as follows:
\begin{enumerate}
  \item Novel Transmission Method : A method that jointly optimize dynamic buffer delay with HARQ to improve QoE in lossy networks.
  \item QoE-Driven Modeling and Analysis:  Firstly, a comprehensive QoE model that integrates key metrics, and the QoE maximization  problem are present. Secondly, the closed-form of the recovery rate and the QoE expression are derived  through rigorous mathematical modeling and analysis. Finally, the optimal transmission parameters are derived. 
  \item Performance Validation: A series of experiments are conducted, the theoretical correctness and effectiveness of the method is validated. 
\end{enumerate}

The rest of this paper is organized as follows. Section II describes the system model and provides a detailed analysis of the QoE model and the maximization  problem, including the derivation of the closed-form expressions and the optimization process.  Experimental results are presented in Section III. Conclusion is drawn in Section IV.

\section{Proposed Method}

\subsection{System Model}

\begin{figure}[t]
  \centering
  \includegraphics[width=1.0\linewidth]{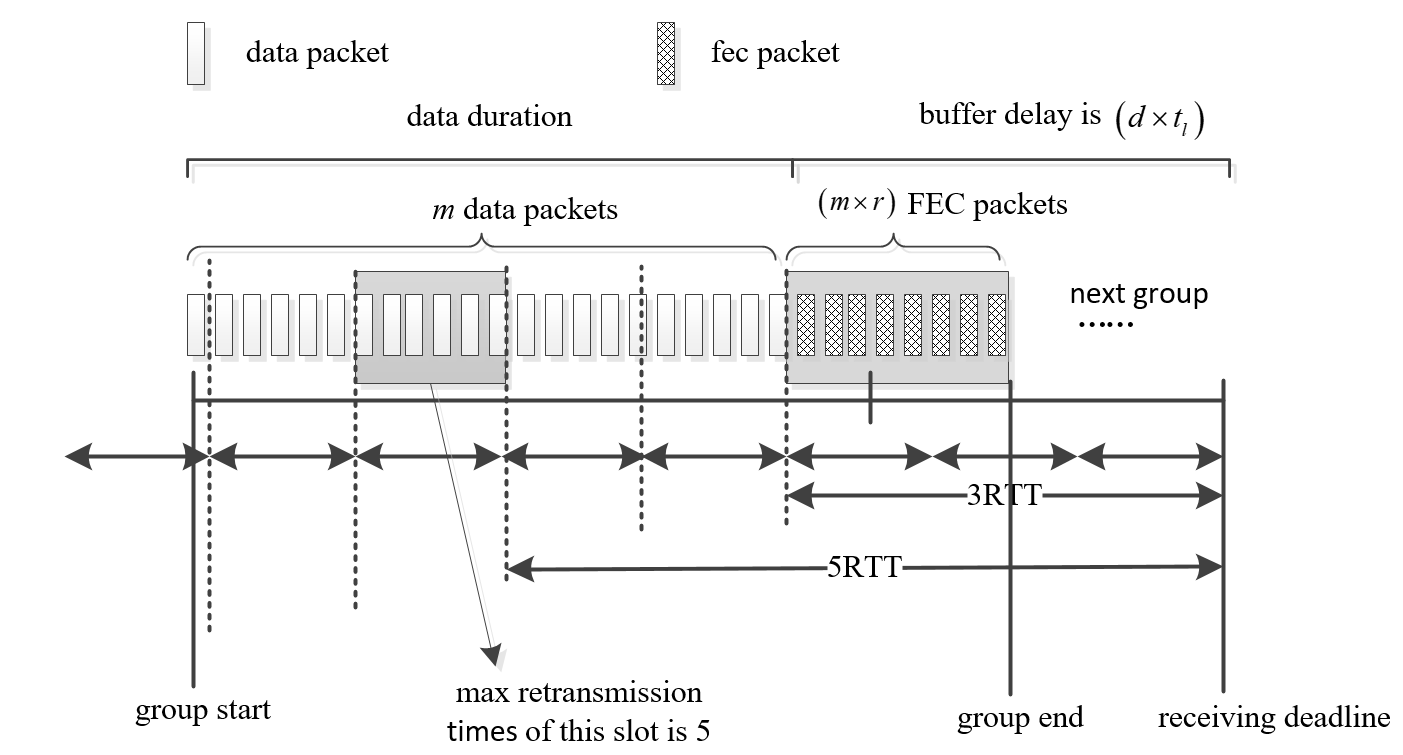}  
  \caption{System Model of the BD-HARQ Method}
  \label{fig:r3transsys}
\end{figure}

The system model for proposed BD-HARQ is depicted in Figure \ref{fig:r3transsys}. Semantic video data, output of the semantic video encoder, is transmitted from the sender to the receiver in groups, with each group consisting of \(m\) data packets and \(m \times r\) FEC packets, where \(r\) represents the packet-level FEC redundancy. The bitrate of semantic data is \(B\), and the size of each packet is \(L\).
The system assumes suffering random packet loss with rate of  \(p\), and the link round trip time (RTT) is \(t_l\). At the receiver, from the time the last data packet is received (i.e., just before the first FEC packet is received) to the packet group receiving deadline (i.e., the time for performing FEC recovery), a dynamic buffer delay with period of \(d \times t_l\) is designed. Additionally,  from the group start up to the receiving deadline, the NACK (Negative Acknowledgement) mechanism \cite{Ahmed2021Hybrid} is used for selective retransmission of lost packets. The probability of successful recovery (i.e., recovery rate) of the packet group at the receiver is denoted as \(\xi\). A successful recovery is only considered when all packets within the group are either received or recovered. This requirement is both reasonable and critical, as the packets within a group typically originate from the same video frame, and the loss of even a single packet can result in the failure to reconstruct the entire frame.

In this system model, the packet loss rate \(p\) and link RTT \(t_l\) are dynamically observed by other system modules. Parameters \(m\) and  \(L\) are preset, \(B\) is determined by the upstream semantic video encoder. Parameters \(d\) and \(r\) are to be optimized. \(\xi\) is determined by network status and transmission parameters, and is to be further analyzed. 

\subsection{QoE Definition and Problem Formulation}

In accordance with the system model, buffer delay \(d\), redundancy \(r\), and packet group recovery rate \(\xi\) are identified as critical factors influencing the QoE. Buffer delay reflects the communication latency, redundancy represents the transmission efficiency, and \(\xi\) indicates the success rate of data recovery, which directly impacts the video quality and playback smoothness. The QoE is defined as:
\begin{equation}\label{eq:QOEt}
    Q = h_{D}Q_{D} + h_{R}Q_{R} + h_{\Xi}Q_{\Xi}
\end{equation}
where \(h_D\), \(h_R\), and \(h_\Xi\) denote the weights assigned to buffer delay, redundancy, and recovery rate,  respectively, subject to the constraint \(h_D + h_R + h_\Xi = 1\). These weights are adjustable to reflect user preferences across different application scenarios. \(Q_D\), \(Q_R\) and \(Q_\Xi\) are the evaluated scores for buffer delay , redundancy, and recovery rate respectively, and are defined as follows:
\begin{equation}\label{eq:fd}
Q_D = \left\{
\begin{array}{ll}
-k_{D,1} \cdot d + 1& \text{if } 0 \leq d \leq d_{1} \\
-k_{D,2} \cdot (d - d_{1}) + f_{D}(d_{1}) & \text{if } d_{1} < d \leq d_{2} \\
-k_{D,3} \cdot (d - d_{2}) + f_{D}(d_{2}) & \text{if } d_{2} < d \leq d_{3} \\
0 & \text{otherwise}
\end{array}
\right.
\end{equation}
\begin{equation}\label{eq:fr}
Q_R = \left\{
\begin{array}{ll}
-k_{R,1} \cdot  r +1 & \text{if } 0 \leq r \leq r_{1} \\
0 & \text{otherwise}
\end{array}
\right.
\end{equation}
\begin{equation}\label{eq:fz}
Q_{\Xi} = \left\{
\begin{array}{ll}
k_{\Xi,1} \cdot \xi & \text{if } 0 \leq \xi \leq \xi_{1} \\
k_{\Xi,2} \cdot (\xi - \xi_{1}) + f_{\Xi}(\xi_{1}) & \text{if } \xi_{1} < \xi \leq \xi_{2} \\
k_{\Xi,3} \cdot (\xi - \xi_{2}) + f_{\Xi}(\xi_{2}) & \text{if } \xi_{2} < \xi \leq 1 \\
0 & \text{otherwise}
\end{array}
\right.
\end{equation}
where the constants \(k_{D,1}\), \(k_{D,2}\), \(k_{D,3}\), \(k_{R,1}\), \(k_{\Xi,1}\), \(k_{\Xi,2}\), and \(k_{\Xi,3}\) are positive values representing the slopes of the respective evaluation functions across different segments, and \(d_{1}\), \(d_{2}\), \(d_{3}\), \(r_{1}\), \(\xi_{1}\), and \(\xi_{2}\) are the boundaries of these segments. The values of \(Q_{D}\), \(Q_{R}\), and \(Q_{\Xi}\) all fall within the range \([0,1]\).

This piecewise linear configuration captures the non-linear characteristics of user perception. Higher slopes are employed in the intermediate stages to indicate increased user sensitivity to changes, while lower slopes are used in less sensitive stages, simulating the user's reduced responsiveness to changes in these intervals.  This approach enables a practical modeling of user reactions under various conditions, and achieving a good balance between accuracy  and simplicity.

Based on \eqref{eq:QOEt}, the problem is formulated as:
\begin{equation}\label{eq:max-qoet}
    \begin{matrix}
    \begin{matrix}
    \max_{r,d} & Q = h_{D}Q_{D} + h_{R}Q_{R} + h_{\Xi}Q_{\Xi} \\
    s.t. & r{\leq r}_{\max} \\
     & d{\leq d}_{\max}
    \end{matrix}
    \end{matrix}
\end{equation}
The optimization aims to determine the optimal values of buffer delay \( d \), redundancy \( r \), and the resulting recovery probability \( \xi \) that maximize the system QoE \( Q \) under the constraints of maximum redundancy \( r_{\max} \) and maximum buffer delay \( d_{\max} \).

\subsection{Optimal Transmission Parameters Derivation}

To address the optimization problem \eqref{eq:max-qoet}, it is crucial to analyze the mathematical relationships between the recovery rate $\xi$, redundancy $r$, and buffer delay $d$. The following sections will systematically discuss: the conditions under which a packet group can be recovered, the probability distribution of the number of lost data packets after retransmission, the probability distribution of the number of lost FEC packets, and finally the recovery rate of the packet group.

\subsubsection{Packet Group Recovery Conditions}

This system employs the RS FEC coding method. Let $S_{m}$, $S_{r}$, and $S_{a}$ denote the number of  respectively data packets, FEC packets, and total packets received for the packet group at the receiving deadline. The packet group can be recovered when the total number of received packets is not less than the number of data packets sent:
\begin{equation}\label{eq:RS-cond}
    S_{a} = S_{m} + S_{r} \geq m
\end{equation}
Alternatively, let $X_{m}$, $X_{r}$, and $X_{a}$ represent the number of lost data packets, lost FEC packets, and total lost packets, respectively. The packet group can be recovered when the total number of lost packets $X_{a}$ is less than the number of FEC packets:
\begin{equation}\label{eq:RS-cond-eq}
    X_{a} = X_{m} + X_{r} \leq mr
\end{equation}

\subsubsection{Probability Distribution of Lost data packets}

To analyze the number and distribution of data packets that remain lost after the retransmission mechanism, the packet group is divided into RTT slots of length $t_{l}$ from right to left. It can be easily derived that the total number of RTT slots for data packets is $N = \left\lfloor \frac{m \times \frac{L}{B} + d \times t_{l}}{t_{l}} \right\rfloor$; the number of data packets contained in each RTT slot is $M = \left\lfloor \frac{t_{l}}{T_{0}} \right\rfloor$, where $T_{0}$ is the interval of data packets, and $T_{0} = \frac{L}{B}$. Here, $\left\lfloor \cdot \right\rfloor$ denotes the floor function.
With the packet loss rate $p$ of the network, under the NACK mechanism, the sender can detect the loss of any data packet and retransmit it after time of $t_{l}$ from last transmission. The maximum retransmission (including the first transmission) times for any data packet in the $j$-th slot is $(j + d)$, and the probability that any data packet still fails to be received is $p_{sj}$:
\begin{equation}\label{eq:psj}
    p_{sj} = p^{j + d}
\end{equation}

Since each slot contains $M$ identically distributed data packets, the number of packets that fail to be received for the $j$-th slot, $X_{sj}$, follows a binomial distribution $X_{sj} \sim B\left( M, p_{sj} \right)$, with mean $\mu_{mj}$ and variance $\sigma_{mj}^{2}$:
\begin{equation}\label{eq:Dis-Data_loss_j}
    \left\{ \begin{matrix}
    \mu_{mj} = Mp_{sj} = Mp^{j + d} \\
    \sigma_{mj}^{2} = Mp_{sj}\left( 1 - p_{sj} \right) = Mp^{j + d}\left( 1 - p^{j + d} \right)
    \end{matrix} \right.\
\end{equation}

Since each packet group contains $N$ independent slots, the number of data packets that fail to be received in the entire packet group, $X_{m}$, is the sum of multiple binomial distributions, with mean $\mu_{m}$ and variance $\sigma_{m}^{2}$:
\begin{equation}\label{eq:Dis-Data_loss}
    \left\{ \begin{matrix}
    \mu_{m} = \sum_{j = 1}^{N}{Mp^{j + d}} \\
    \sigma_{m}^{2} = \sum_{j = 1}^{N}{Mp^{j + d}\left( 1 - p^{j + d} \right)}
    \end{matrix} \right.\
\end{equation}

\subsubsection{Probability Distribution of Lost FEC Packets}

Given redundancy $r$, the number of FEC packets is $mr$. After transmission, the number of lost FEC packets $X_{r}$ follows a binomial distribution: $X_{r} \sim B(mr, p)$. The mean $\mu_{r}$ and variance $\sigma_{r}^{2}$ are given by:
\begin{equation}\label{eq:Dis-Red_loss}
    \left\{ \begin{matrix}
    \mu_{r} = mrp \\
    \sigma_{r}^{2} = mrp(1 - p)
    \end{matrix} \right.\
\end{equation}

\subsubsection{Probability Distribution of Total Lost Packets}

Since the transmission processes of data packets and FEC packets are independent, the total number of lost packets $X_{a}$ at the receiving deadline is the sum of the number of lost data packets $X_{m}$ and the number of lost FEC packets $X_{r}$. Therefore, the mean $\mu_{a}$ and variance $\sigma_{a}^{2}$ of the total lost packet are:
\begin{equation}\label{eq:Dis-All_loss}
    \left\{ \begin{matrix}
    \mu_{a} = \mu_{m} + \mu_{r} = \sum_{j = 1}^{N}{Mp^{j + d}} + mrp \\
    \sigma_{a}^{2} = \sigma_{m}^{2} + \sigma_{r}^{2} = \sum_{j = 1}^{N}{Mp^{j + d}\left( 1 - p^{j + d} \right)} + mrp(1 - p)
    \end{matrix} \right.\
\end{equation}
It is evident that the number of total lost packets $X_{a}$ is the sum of multiple binomial distributions, and its mean, variance, and probability distribution are functions of buffer delay $d$ and redundancy $r$. In practice, $X_{a}$ can be further approximated by a Gaussian distribution:
\begin{equation}\label{eq:Dis-All_loss-norm}
    X_{a} \sim N\left( \mu_{a}, \sigma_{a}^{2} \right)
\end{equation}

\subsubsection{Recovery Ratio of the Packet Group}

By combining the previously derived condition of successful recovery for the packet group in \eqref{eq:RS-cond-eq} with the distribution of the total number of lost packets $X_{a}$ in \eqref{eq:Dis-All_loss-norm}, the probability that the packet group can successfully recover is:
\begin{equation}\label{eq:p-recover}
\begin{matrix}
    \xi = P\left( X_{a} \leq mr \right) = \Phi\left( \frac{mr - \mu_{a}}{\sigma_{a}} \right) \\ 
    = \Phi\left( \frac{mr - \left( \sum_{j = 1}^{N}{Mp^{j + d}} + mrp \right)}{\sqrt{\sum_{j = 1}^{N}{Mp^{j + d}\left( 1 - p^{j + d} \right)} + mrp(1 - p)}} \right)
 \end{matrix}
\end{equation}
where $\Phi(\cdot)$ is the cumulative distribution function of the standard normal distribution, typically computed using lookup tables.

It is clear that, when the network state parameters $p$ and $t_{l}$ are known, for semantic data with bitrate $B$, under the preset parameters $m$ and $L$, the recovery rate $\xi$ is determined by the redundancy $r$ and the buffer delay $d$.

\subsubsection{Problem Solving}

By integrating \eqref{eq:fd}, \eqref{eq:fr}, \eqref{eq:fz} and \eqref{eq:p-recover} with \eqref{eq:QOEt}, it becomes apparent that for problem \eqref{eq:max-qoet},  $Q_{D}$ and $Q_{R}$ are piecewise linear functions with respect to the optimization variables $r$ and $d$, and $Q_{\Xi}$ is a piecewise linear function of $\xi$, which further translates into a piecewise linear mapping function of the Gaussian cumulative distribution function concerning $r$ and $d$. The constraints are straightforward boundary conditions on $r$ and $d$.

In practice, discretization constraints can be further imposed on \(r\) and \(d\), for example, with step sizes of \(1/m\) and \(0.5t_l\), respectively. Consequently, the discrete feasible solutions for $r$ and $d$ are constrained within a limited space, allowing the optimization problem \eqref{eq:max-qoet} to be solved using an exhaustive search at minimal computational cost.

\section{Experiments and Results Analysis}

To validate the effectiveness of the proposed method, a series of experiments were conducted in a controlled laboratory environment. The network impairments were emulated using the hardware device Spirent Attero, where the sender transmitted packets to the receiver over the lossy network. All packet transmissions utilized the Real-time Transport Protocol (RTP).

\subsection{Detailed Experimental Parameters}

The system parameters applied in the QoE  model and the BD-HARQ method in the experiments are listed in Table ~\ref{tab:qoe_params} and  ~\ref{tab:hqarq_params} , providing details on the parameters , including their names, meanings, specific values, and units (with all time units normalized to 100ms):

\begin{table*}[tb]
  \centering
  \caption{QoE Model Parameters}
    \begin{tabular}{p{2.5cm}p{4.0cm}p{2.5cm}p{2.5cm}}
    \hline
    Parameter & Meaning & Value & Unit \\
    \hline
    $h_D$ & Buffer Delay Indicator Weight & 0.3 & - \\
    $h_R$ & Redundancy Indicator Weight & 0.3 & - \\
    $h_\Xi$ & Semantic Data Recovery Rate Indicator Weight& 0.4 & - \\
    $k_{D,1}$, $k_{D,2}$, $k_{D,3}$ & Buffer Delay Evaluation Function Slopes & 0.04, 0.32, 0.07 & 1/(100ms) \\
    $d_1$, $d_2$, $d_3$ & Buffer Delay Evaluation Function Segmentation Points & 2.5, 8.0, 31.0 & 100ms \\
    $k_{R,1}$ & Redundancy Evaluation Function Slope & 2 & - \\
    $r_1$ & Redundancy Evaluation Function Segmentation Point & 0.5 & - \\
    $k_{\Xi,1}$, $k_{\Xi,2}$, $k_{\Xi,3}$ & Semantic Data Recovery Rate Evaluation Function Slopes & 0.40, 1.67, 1.00 & - \\
    $\xi_1$, $\xi_2$ & Semantic Data Recovery Rate Evaluation Function Segmentation Points & 0.5, 0.95 & - \\
    \hline
    \end{tabular}
  \label{tab:qoe_params}
\end{table*}

\begin{table*}[tb]
  \centering
  \caption{Preset System Parameters}
    \begin{tabular}{p{2.5cm}p{4.0cm}p{2.5cm}p{2.5cm}}
    \hline
    Parameter & Meaning & Value & Unit \\
    \hline
    $t_l$ & Network Link Round Trip Time & 1 & 100ms \\
    $m$ & Number of Data Packets per Group& 16 & - \\
    $B$ & Bitrate of Data Packets& 1000 & Kbps \\
    $L$ & Size of Each Data Packet & 1000 & Bytes \\
    $r_{\max}$ & Maximum Redundancy & 0.5 & - \\
    $d_{\max}$ & Maximum Buffer Delay& 10.0 & 100ms \\
    \hline
    \end{tabular}
  \label{tab:hqarq_params}
\end{table*}

\begin{figure}[tbp]
  \centering
  \includegraphics[width=1.0\linewidth]{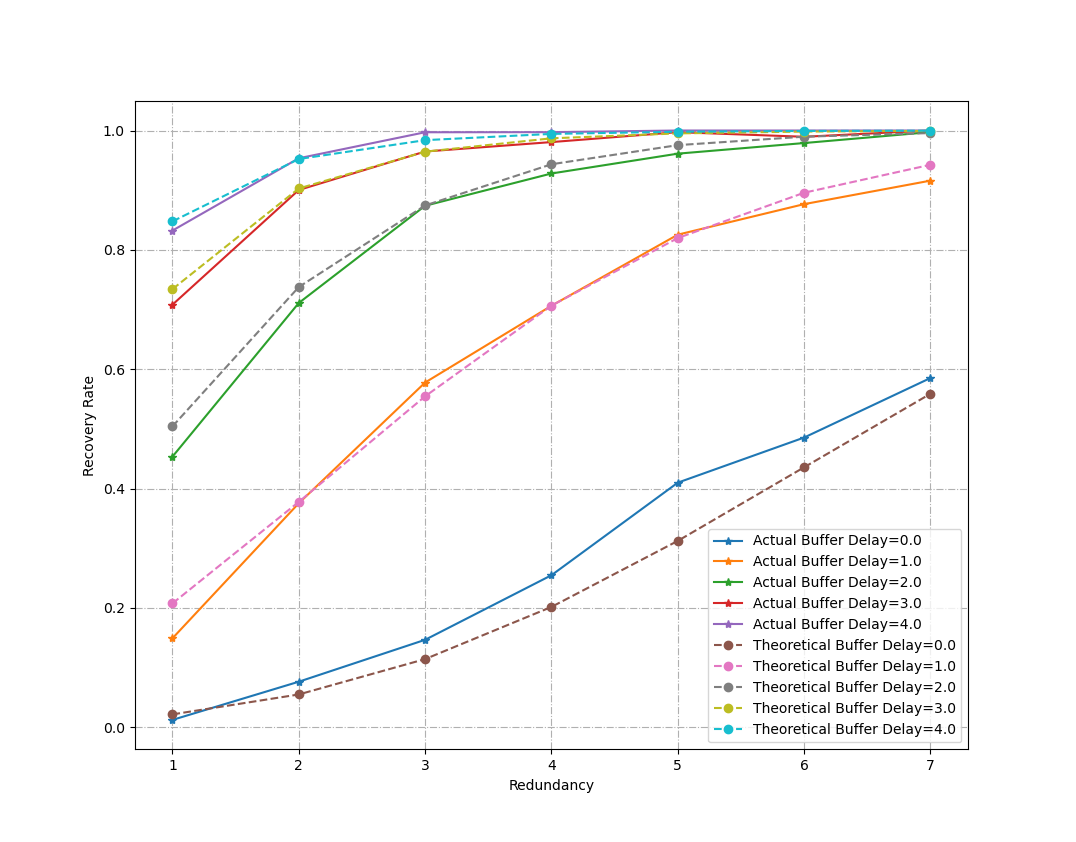}  
  \caption{Relationship Between Recovery Rate and Redundancy Under Different Buffer Delays}
  \label{fig:r3rp}
\end{figure}

\begin{figure}[tbp]
  \centering
  \includegraphics[width=1\linewidth]{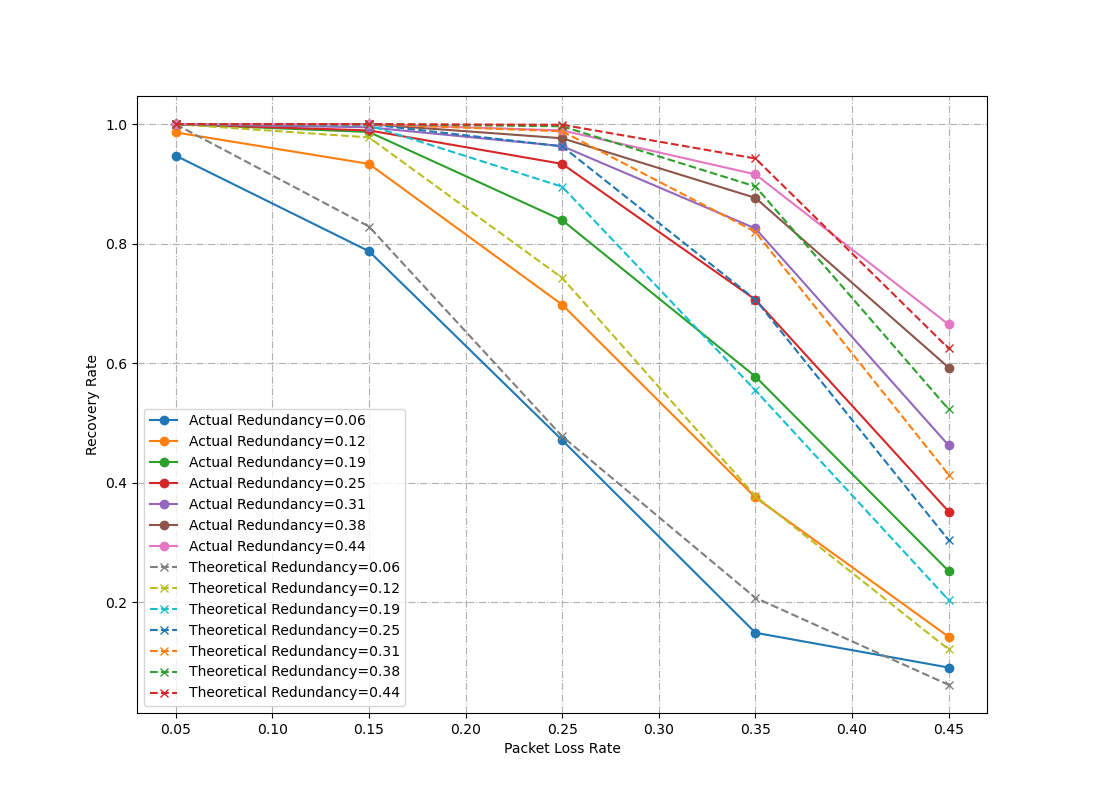}  
  \caption{Relationship Between Recovery Rate and Packet Loss Rate Under Different Redundancies}
  \label{fig:r3pp}
\end{figure}

\subsection{Experimental Results Analysis}

Fig.~\ref{fig:r3rp} and~\ref{fig:r3pp} present the variations of recovery rate with different network status and transmission parameters, comparing experimental results with theoretical values.

Fig.~\ref{fig:r3rp} illustrates the trend of the system recovery rate with redundancy and buffer delay when the packet loss rate $p$ is 0.35. It is evident that under various combinations of buffer delay and redundancy,  all experimental and theoretical recovery rates match well, validating the correctness of the theoretical analysis. The trend indicates that, given a certain buffer delay, the recovery rate improves with increased redundancy, but the pace of improvement diminishes when \(\xi\) exceeds about 0.6. Similarly, with fixed redundancy, the recovery rate increases as the buffer delay grows, though this improvement also tapers off when \(\xi\) surpasses 0.6. The figure further demonstrates a negative correlation between buffer delay and redundancy for achieving a certain recovery rate: a higher redundancy necessitates a lower buffer delay, and vice versa.

Fig.~\ref{fig:r3pp} shows the trend of the system recovery rate with the network packet loss rate and redundancy when the buffer delay $d=2.0$, i.e., 200ms. As observed in Fig.~\ref{fig:r3rp}, the recovery rates from experimental and theoretical values math well under various combinations of packet loss rates and redundancies, further validating the theoretical analysis. The trend reveals that with certain redundancy, the recovery rate decreases as the packet loss rate increases, initially rapidly, then more gradually. Similarly, with a certain packet loss rate, the recovery rate increases with increased redundancy, following a trend similar to that in Fig.~\ref{fig:r3rp}.

\begin{figure}[tbp]
  \centering
  \includegraphics[width=1\linewidth]{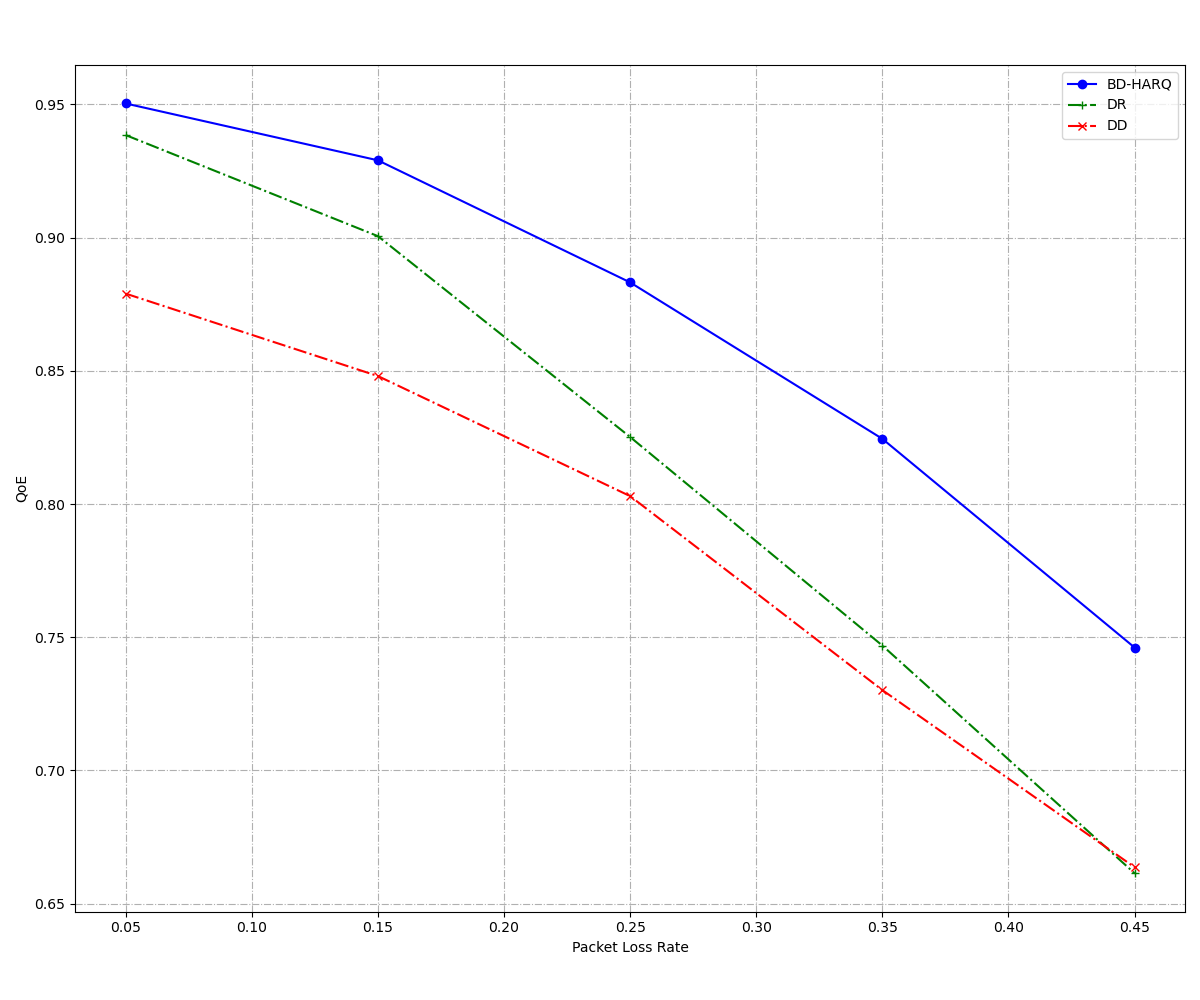}  
  \caption{Comparison of QoE for Different Transmission Methods}
  \label{fig:r3pq}
\end{figure}

Fig.~\ref{fig:r3pq} compares the QoE performance of the proposed BD-HARQ method with two commonly used alternatives, the DR and DD methods, under varying packet loss rates. The DR method employs dynamic redundancy with fixed buffer delay, while the DD method utilizes dynamic buffer delay with fixed redundancy. 
The results indicate that while the QoE of all transmission methods declines as packet loss rate increases, the proposed BD-HARQ method consistently outperforms the alternatives. At low packet loss rates, the DR method's performance approaches that of BD-HARQ, whereas the DD method lags significantly. As packet loss rate rises, the QoE of the DR method deteriorates more rapidly, eventually converging with that of the DD method, both falling significantly below the BD-HARQ method.  For instance, at $p=0.05$,  the QoE of the proposed method improves from 0.94 and 0.87 for the DR and DD methods to 0.95, representing an increase of 1.1\% and 9.2\%, respectively. At $p=0.35$, the proposed method enhances QoE from 0.75 and 0.73 for the DR and DD methods to 0.83, reflecting improvements of 10.7\% and 13.7\%, respectively. Furthermore, to maintain a certain QoE of 0.85, the proposed method increases the tolerance for packet loss rate from the original 15\%-22\% to as much as 31\%, demonstrating a significant enhancement in robustness against network conditions.

\begin{figure}[tbp]
  \centering
  \includegraphics[width=1\linewidth]{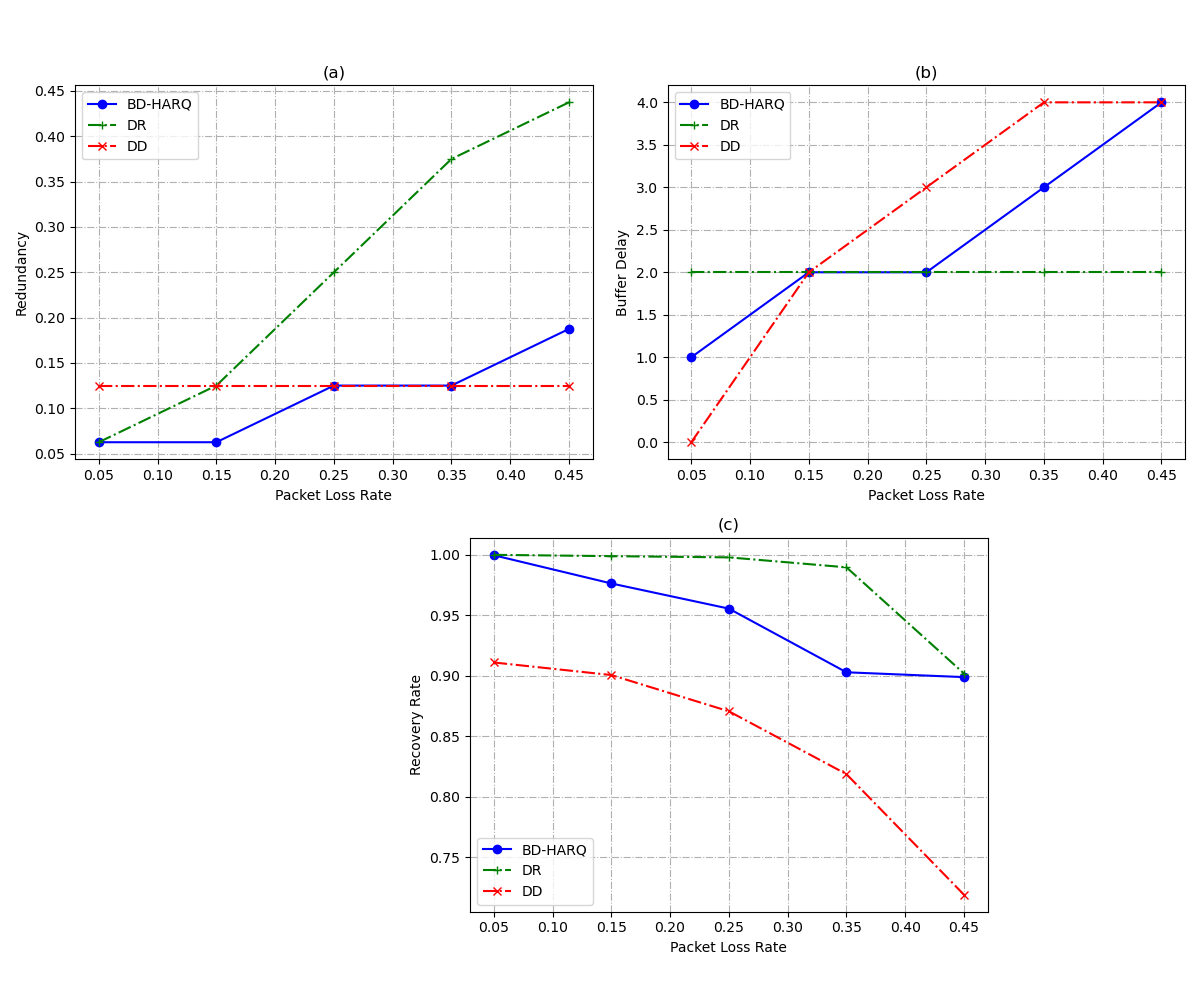}  
  \caption{Comparison of Detailed Metrics for Different Transmission Methods}
  \label{fig:r3com-abc}
\end{figure}

To further analyze the QoE differences between the methods depicted in Fig.~\ref{fig:r3pq}, a detailed comparison of redundancy, buffer delay, and recovery rate at different packet loss rates is provided, as shown in Fig.~\ref{fig:r3com-abc}.

From Fig.~\ref{fig:r3com-abc}(a), it is evident that the proposed BD-HAQR method generally employs a lower redundancy setting. At low packet loss rates, it utilizes lower redundancy, which increases gradually as the packet loss rate rises, and surpasses the DD method at a high packet loss rate of $p=0.45$. The DR method's redundancy increases significantly with the packet loss rate, exceeding that of the proposed BD-HAQR and DD methods.

Fig.~\ref{fig:r3com-abc}(b) indicates that the BD-HARQ method adopts a moderate buffer delay setting. At the same packet loss rate, its buffer delay setting is generally intermediate between those of the DR and DD methods. At low packet loss rates, the BD-HARQ method's buffer delay is lower than that of the DR method but higher than that of the DD method. As the packet loss rate increases, both the BD-HARQ and DD methods' buffer delays gradually increase, with the BD-HARQ method maintaining a lower buffer delay than the DD method.

Comparing the redundancy and buffer delay of different methods in ~\ref{fig:r3com-abc}(a) and ~\ref{fig:r3com-abc}(b) with the recovery rates in Fig.~\ref{fig:r3com-abc}(c), it is clear that under respective transmission parameters,  the DR method achieves the highest recovery rate, and that the proposed BD-HAQR method also maintains a high recovery rate that declines gradually with an increasing packet loss rate. The DD method exhibits the lowest recovery rate, which declines significantly with an increasing packet loss rate.

Comparing the DR method with the proposed BD-HAQR method, although the former achieves advantageous recover rate, its final QoE is lower than that of the BD-HAQR method. This indicates that although the fixed buffer delay and significantly increased redundancy strategy can improve the recovery rate, the advantage in recovery rate is insufficient to offset the overall QoE degradation caused by the substantial increase in redundancy. Comparatively, the DD method, with its fixed redundancy, performs poorly in both recovery rate and buffer delay, resulting in lower QoE than the proposed BD-HAQR method.

\section{Conclusion}

This paper studies the issue of video transmission in lossy network environments and proposes a dynamic buffer delay and hybrid automatic repeat request joint optimization method. Based on the establishment of an indicator system that comprehensively considers user experience quality, including buffer delay, redundancy, and semantic data recovery rate, this paper reveals the intrinsic connections between these indicators through mathematical modeling and in-depth analysis, and dynamically adjusts the buffer delay and redundancy. Experimental results validate the theoretical correctness and effectiveness of the BD-HARQ method in improving user QoE. In future, researches on real-time accurate estimation of network status and integration of cross-layer design will be explored, expecting  better system performance and QoE.

% \section*{References}

\vspace{12pt}
\color{red}

\end{document}